# GEOMETRIC ORIGIN OF PHYSICAL CONSTANTS IN A KALUZA-KLEIN TETRAD MODEL

Frank Reifler and Randall Morris

Lockheed Martin Corporation MS2 137-205

199 Borton Landing Road

Moorestown, NJ 08057

ABSTRACT: An important feature of Kaluza-Klein theories is their ability to relate fundamental physical constants to the radii of higher dimensions. In previous Kaluza-Klein theory, which unifies the electromagnetic field with gravity as dimensionless components of a Kaluza-Klein metric, i) all fields have the same physical dimensions, ii) the Lagrangian has no explicit dependence on any physical constants except mass, and hence iii) all physical constants in the field equations except for mass originate from geometry. While it seems natural in Kaluza-Klein theory to add fermion fields by defining higher dimensional bispinor fields on the Kaluza-Klein manifold, these Kaluza-Klein theories do not satisfy conditions (i), (ii), and (iii). In this paper, we show how conditions (i), (ii), and (iii) can be satisfied by including bispinor fields in a tetrad formulation of the Kaluza-Klein model, as well as in an equivalent teleparallel model. This demonstrates an unexpected feature of Dirac's bispinor equation, since conditions (i), (ii), (iii) imply a special relation among the terms in the Kaluza-Klein or teleparallel Lagrangian that would not be satisfied in general.

## 1. INTRODUCTION

An important feature of Kaluza-Klein theories is their ability to relate fundamental physical parameters such as Newton's gravitational constant and the charges associated with gauge fields to the radii of higher dimensions [1] – [6]. In previous Kaluza-Klein theory, which unifies gravity with electromagnetism, the formula, $q = \sqrt{16\pi\kappa_0}/\hat{\delta}$, allows the electric charge $q$ to be expressed as a function of Newton's constant $\kappa_0$ and the radius $\hat{\delta}$ of an extra dimension [2], [3], [6]. (Note that with the speed of light c and Planck's constant $\hbar$ set equal to one, $\sqrt{\kappa_0}$ and $\hat{\delta}$ have units of length.) This formula requires Newton's constant to define the electric charge in terms of the higher dimensional radius. Nevertheless, it can be easily shown that by unifying the electromagnetic field with gravity as dimensionless components of a Kaluza-Klein metric, both Newton's constant and the electric charge can be replaced in the Kaluza-Klein equations by the higher dimensional radius $\hat{\delta}$ (see Section 2).

A reasonable goal for a Kaluza-Klein model is that all fields have the same physical dimensions, as well as that all physical constants should originate from geometry in a unified field theory. With this in mind, we define a "geometric model" to be one in which all fields (like the gravitational field) are dimensionless, and with no physical constants (except mass) appearing explicitly in the Lagrangian. (It is assumed in this definition that no special choices of units are made beyond setting the speed of light c and Planck's constant $\hbar$ equal to one.) It is natural in Kaluza-Klein theory to add fermion fields by simply defining higher dimensional bispinor fields on the Kaluza-Klein manifold [2], [7], [8]. These Kaluza-Klein theories, however, do not satisfy the two conditions for a "geometric model" just stated. In this paper, we show that these conditions can be satisfied by including bispinor fields in a tetrad formulation of the Kaluza-Klein model [9] – [12]. This unexpected feature of Dirac's bispinor equation is nontrivial, since we show in Sections 2 and 3 that the two conditions taken together imply a special relation among the terms in the Kaluza-Klein Lagrangian that would not be satisfied in general.

For example, simple rescaling of the fields will not eliminate Newton's constant from the usual Einstein-Maxwell-Dirac Lagrangian when all interaction terms are included. That is, assuming a unit electric charge, the interaction term $V_\alpha j^\alpha$ is linear in the electromagnetic gauge potential $V_\alpha$ and the fermion current $j^\alpha$, and hence quadratic in the bispinor field $\Psi$. Also, there are non-interacting terms quadratic in $V_\alpha$ and quadratic in $\Psi$. Thus, Newton's constant cannot be eliminated from the usual Einstein-Maxwell-Dirac Lagrangian by simply rescaling $V_\alpha$ and $\Psi$. As another example, consider a Kaluza-Klein tetrad model in which the Kaluza-Klein metric depends on a tetrad and a scalar field [9]. For this case, as well as its teleparallel equivalent, the elimination of Newton's constant along with all coupling constants is possible only when a special relation holds between various terms in the Lagrangian (see Sections 2 and 3).

The Kaluza-Klein tetrad model presented here requires mapping bispinor fields into their tensor equivalents. We shall briefly review this map to clarify the technical development in the following sections. Using geometric algebra, Hestenes showed in 1967 that a bispinor field on a Minkowski space-time is equivalent to an orthonormal tetrad of vector fields together with a complex scalar field, and that fermion plane waves can be represented as rotational modes of the tetrad [13]. As stated by Takahashi, who in 1983 derived the tensor form of the Dirac bispinor Lagrangian in terms of Hestenes' tetrad and scalar field, and who is well known for his work in quantum field theory [14] – [16], "…. a tetrad in a Minkowski space implies the existence of a spinor, and the [spinor] orthogonality and completeness conditions are automatically satisfied, when the tetrad is expressed in terms of the spinor." [14] These orthogonality and completeness conditions give rise to the oscillator modes that lead directly to fermion creation and annihilation operators [17]. Indeed, the oscillator modes of a bispinor field are precisely the oscillator modes of a tetrad field. From this point of view, as recognized by both Hestenes and Takahashi, there can be no difference between the bispinor and tensor theories for establishing the quantum field theory of fermions.

However, there are two assumptions which underlie Takahashi's claim that Hestenes' tensor fields are equivalent to bispinors and that they describe fermions. The first assumption is that global tetrad fields exist on the space-time. This assumption is satisfied if the space-time is four-dimensional, admits spinor structure, and is non-compact [18] – [23]. On such space-times, spinor structures and homotopy classes of global tetrad fields are synonymous [20]. The second assumption, required for a consistent interpretation within quantum mechanics, is that we restrict to solutions of the tensor Dirac equation having "unique continuation" [17], [24], [25]. That is, for any observer, the history of a tetrad field in the past must uniquely determine its evolution into the future [24] – [29]. Non-unique continuation of solutions occurs in other areas of physics, such as in fluid dynamics, but is not considered to be appropriate for quantum mechanics. Note that there can be a bispinor field having unique continuation, whose tetrad field cannot be uniquely predicted into the future [17], [26], [27]. However, in a Minkowski space-time, continuation of the tetrad field is unique if we restrict to physically realizable bispinor solutions of the Dirac equation whose energy spectrum is bounded from below [17], [24], [25], [28], [29]. For example, the thought experiment [26] proposed by Y. Aharonov and L. Susskind to observe the sign change of bispinors under $2\pi$ rotations is not physically realizable, because the bispinor field would have to have an unbounded energy spectrum stretching from negative infinity to positive infinity. This could only be achieved by superposition of particle and anti-particle solutions, which is forbidden by a superselection rule in quantum mechanics [17].

In this paper, we treat a bispinor field $\Psi$ as a classical field, rather than as a quantum mechanical wave function as in the discussion above. It has been shown that every bundle of spin frames on a non-compact four-dimensional space-time with spinor structure is trivial [20], [23]. Hence, on any open subset U of the space-time where a reference tetrad is defined, bispinor fields $\Psi$ can be simply expressed as maps $\Psi : U \to C^4$, where $C^4$ is a four-dimensional complex vector space [20]. Furthermore, in a non-compact four-dimensional space-time with spinor structure, the Einstein-Dirac equations depend only on a tetrad and a scalar field [30], [31]. This can be demonstrated by an appropriate choice of reference tetrad [12]. The appropriate choice is provided by Hestenes' orthonormal tetrad of vector fields, denoted as $e_a^\alpha$, where $\alpha = 0, 1, 2, 3$ is a space-time index and $a = 0, 1, 2, 3$ is a tetrad index [13]. Relative to this special reference tetrad, a bispinor field $\Psi$ is "at rest" at each space-time point and has components given as follows [12]:

$$\Psi = \begin{bmatrix} 0 \\ \mathrm{Re}[\sqrt{s}] \\ 0 \\ -i\,\mathrm{Im}[\sqrt{s}] \end{bmatrix} \qquad (1.1)$$

where s is a complex scalar field [12], [13]. Note that Hestenes' tetrad $e_a^\alpha$ and the complex scalar field $\sqrt{s}$ are smoothly defined locally in open regions about each space-time point where s is nonvanishing. In each open region, the Dirac bispinor Lagrangian depends only on the reference tetrad $e_a^\alpha$ and on the bispinor field $\Psi$. [32] – [38] It was shown previously [12], using formula (1.1), that the Einstein-Dirac Lagrangian can be expressed entirely in terms of the tensor fields, $e_a^\alpha$ and s, once Hestenes' tetrad has been chosen as the reference.

Whenever $\Psi$ vanishes, both s and its first partial derivatives vanish. Setting s and its first partial derivatives to zero in the tensor form of Dirac's bispinor equation shows that $e_a^\alpha$ can be chosen arbitrarily at all space-time points where $\Psi$ vanishes. Thus, all aspects of Dirac's bispinor equation are faithfully reflected in the tensor equation [12]. Since the tetrad $e_a^\alpha$ is unconstrained by the Dirac equation when $\Psi$ vanishes, a gravitational field exists even if the fermion field vanishes. Thus, the gravitational field $g_{\alpha\beta}$ and the bispinor field $\Psi$ (which together have 10 + 8 = 18 real components), are represented accurately by Hestenes' tensor fields $e_a^\alpha$ and s (which also have 16 + 2 = 18 real components) [9] – [12].

The Kaluza-Klein tetrad model is based on a constrained Yang-Mills formulation of the Dirac theory [9] – [12]. In this formulation Hestenes' tensor fields $(e_a^\alpha, s)$ are mapped bijectively onto a set of $SL(2,R) \times U(1)$ gauge potentials $F_\alpha^K$ and a complex scalar field $\rho$. Thus we have the composite map $\Psi \to (e_a^\alpha, s) \to (F_\alpha^K, \rho)$. The fact that $e_a^\alpha$ is an orthonormal tetrad of vector fields imposes an orthogonal constraint on the gauge potentials $F_\alpha^K$ given by [12]:

$$F_\alpha^K F_{K\beta} = |\rho|^2 g_{\alpha\beta} \qquad (1.2)$$

where $g_{\alpha\beta}$ denotes the space-time metric. The gauge index K = 0, 1, 2, 3 is lowered and raised using a gauge metric $g_{JK}$ and its inverse $g^{JK}$ (see Section 2). Repeated indices are summed. It was shown previously that via the map $\Psi \to (F_\alpha^K, \rho)$ the Dirac bispinor Lagrangian equals a constrained Yang-Mills Lagrangian in the limit of an infinitely large Yang-Mills coupling constant [9] – [12]. In both Lagrangians we may include an electromagnetic gauge field $V_\alpha$ interacting with the fermion field $(F_\alpha^K, \rho)$. It was shown previously [10] that $(F_\alpha^K, V_\alpha)$ are gauge potentials for a Lie group that is a semi-direct product of fermion and electromagnetic gauge groups with self-coupling constants g and q. The Yang-Mills Lagrangian for the semi-direct product gauge potentials $(F_\alpha^K, V_\alpha)$ and the complex scalar field $\rho$ is given by [10], [12]:

$$L_{YM} = -\frac{1}{4} V_{\alpha\beta} V^{\alpha\beta} + \frac{1}{4g} F_{\alpha\beta}^K F_K^{\alpha\beta} + \frac{1}{g_0} \overline{D_\alpha(\rho+\mu)} D^\alpha(\rho+\mu) \qquad (1.3)$$

where $(F_{\alpha\beta}^K, V_{\alpha\beta})$ are the components of the Yang-Mills field tensor associated with $(F_\alpha^K, V_\alpha)$, and $D_\alpha$ is the Yang-Mills covariant derivative acting on the complex scalar field $\rho$ and mass parameter $\mu$. Both $\rho$ and $\mu$ are coupled to the U(1) fermion gauge potential $F_\alpha^3$ with coupling constant $g_0 = \frac{3}{2}g$, and the fermion mass is given by $m_0 = \frac{1}{2}g_0\mu$. In the limit that g becomes infinitely large, the Yang-Mills Lagrangian (1.3) equals precisely Maxwell's Lagrangian for the U(1) electromagnetic gauge potential $V_\alpha$ plus Dirac's Lagrangian for a bispinor field $\Psi$ interacting with $V_\alpha$. Note that both the orthogonal constraint (1.2) and the limit are explicated in a Kaluza-Klein model [9], [10].

As discussed above, the elimination of Yang-Mills coupling constants along with Newton's constant in Kaluza-Klein theory, which includes fermion fields, is a significant, new result. Previously coupling constants were eliminated only with fermion fields excluded from the Kaluza-Klein model. Based on the Yang-Mills Lagrangian (1.3) this elimination is nontrivial because it requires use of the following four conditions:

a) The semi-direct product structure of the fermion and boson gauge potentials
b) The multiple coupling constants $g$, $g_0$, and $q$ for the semi-direct product gauge group
c) The orthogonal constraint, which forces the fermion gauge potentials $F_\alpha^K$ to be related to the scalar field $\rho$ in formula (1.2)
d) The non-standard form of the Yang-Mills Lagrangian in formula (1.3) including the specific coefficients multiplying each term

For example, simply replacing $g_0$ with unity in the Lagrangian (1.3), without introducing a new constant in the orthogonal constraint (1.2), would make it impossible to eliminate explicit dependence of the Kaluza-Klein Lagrangian on Newton's constant and all Yang-Mills coupling constants. We show in Sections 2 and 3 that eliminating explicit dependence of the Lagrangian on these constants in a Kaluza-Klein tetrad model is only possible when the terms are multiplied by specific coefficients as in the Lagrangian (1.3).

## 2. THE ORIGIN OF PHYSICAL CONSTANTS IN THE KALUZA-KLEIN TETRAD MODEL

Let $M = X \times G$ be the Kaluza-Klein manifold, with $X$ a four-dimensional space-time, and $G$ a Lie group. On the space-time X, we assume the existence of a global, nonsingular tetrad of smooth one-forms $\beta^K$ with K = 0, 1, 2, 3. (This assumption simplifies the derivations in this section, but is stronger than necessary, since it is possible to allow tetrads to be singular at exceptional points [12].) The gravitational field on X, which we denote as $\beta$, is defined to be the unique metric tensor with the Minkowski signature, for which the tetrad $\beta^K$ is orthonormal:

$$\beta = g_{JK} \beta^J \otimes \beta^K \qquad (2.1)$$

where

$$g_{JK} = g^{JK} = \begin{bmatrix} 1 & 0 & 0 & 0 \\ 0 & -1 & 0 & 0 \\ 0 & 0 & -1 & 0 \\ 0 & 0 & 0 & -1 \end{bmatrix} \qquad (2.2)$$

Repeated indices J, K, L, M are summed from 0 to 3. The components of $\beta^K$ with respect to local coordinate one-forms $dx^\alpha$ are denoted by $\beta_\alpha^K$. That is, $\beta^K = \beta_\alpha^K dx^\alpha$. Note that $\beta^K$ and $dx^\alpha$ have units of length, whereas the components $\beta_\alpha^K$ are dimensionless. Also note that in formula (2.1) the metric components $g_{JK}$ are dimensionless.

We consider as a Lie group the semi-direct product of the $U(1)$ electromagnetic gauge group with the fermion $SL(2,R) \times U(1)$ gauge group [10]. However, the action of this boson $U(1)$ gauge group on the fermion $SL(2,R) \times U(1)$ gauge group in the semi-direct product is nontrivial, and to simplify, notation will be handled as follows. First, we enlarge the boson $U(1)$ gauge group to an $SL(2,R) \times U(1)$ group by embedding it as a $U(1)$ subgroup of $SL(2,R)$. Then, we consider the semi-direct product of the boson and fermion $SL(2,R) \times U(1)$ gauge groups, with the boson gauge group acting on the fermion gauge group by the adjoint representation [10]. Later, to recover the exact Einstein-Maxwell-Dirac Lagrangian, we will restrict the boson gauge group to just the $U(1)$ electromagnetic gauge group acting on the fermion gauge group.

On this eight-dimensional Lie group, denoted as $G$, we fix two nonsingular tetrads that form a basis of right invariant one-forms $\alpha^K$ and $\hat{\alpha}^K$ with K = 0, 1, 2, 3. The two tetrads of right invariant one-forms $\alpha^K$ and $\hat{\alpha}^K$ define a right invariant metric on the Lie group $G$ given by:

$$\alpha = g_{JK}(\alpha^J \otimes \alpha^K - \hat{\alpha}^J \otimes \hat{\alpha}^K) \qquad (2.3)$$

where $g_{JK}$ is the Minkowski metric in the definition (2.2). The basis of right invariant one-forms $(\alpha^K, \hat{\alpha}^K)$ uniquely determines a dual basis of right invariant vector fields $(a_K, \hat{a}_K)$ on $G$. From formula (2.3) the right invariant vector fields $(a_K, \hat{a}_K)$ form an orthonormal basis for the Lie algebra of $G$. We can choose the basis of right invariant one-forms $\alpha^K$ and $\hat{\alpha}^K$ so that the vector fields $a_K$ and $\hat{a}_K$ satisfy the following semi-direct product commutation relations:

$$[a_J, a_K] = \delta^{-1} f_{JK}^L a_L$$

$$[\hat{a}_J, a_K] = \hat{\delta}^{-1} f_{JK}^L a_L \qquad (2.4)$$

$$[\hat{a}_J, \hat{a}_K] = \hat{\delta}^{-1} f_{JK}^L \hat{a}_L$$

where $\delta$ and $\hat{\delta}$ are length parameters and $f_{JK}^L$ are the Lie algebra structure constants for the Lie group $SL(2,R) \times U(1)$. Note that formula (2.4) leads to a different choice of Lie algebra basis for the boson gauge group than previous work [10]. Also, note that while $\delta$ and $\hat{\delta}$ are the radii of $U(1)$ subgroups of the $SL(2,R)$ factors of $G$, the commutation relations (2.4) do not depend on the radii $\delta_0$ and $\hat{\delta}_0$ of the $U(1)$ factors of $G$. Recall that $G$ has two $U(1)$ factors, generated by the vector fields $a_3$ and $\hat{a}_3$, since $G$ is the semi-direct product of $SL(2,R) \times U(1)$ with itself.

As on any physical manifold, the one-forms $\alpha^K$ and $\hat{\alpha}^K$ carry units of length, so that their dual vector fields $a_K$ and $\hat{a}_K$ carry units of mass (i.e., inverse length). Note that the structure constants $f_{JK}^L$ are dimensionless, so that the length parameters $\delta$ and $\hat{\delta}$ are required in formula (2.4) to balance the dimensions.

Thus on the Kaluza-Klein manifold $M = X \times G$, we can define a dynamic tetrad of one-forms $\beta^K$ and two fixed (constant) tetrads of one-forms $\alpha^K$ and $\hat{\alpha}^K$, induced from the projections of $M = X \times G$ onto its factors $X$ and $G$. ($\beta^K$, $\alpha^K$, and $\hat{\alpha}^K$ on $M = X \times G$ are the pullbacks of $\beta^K$ on $X$, and $\alpha^K$ and $\hat{\alpha}^K$ on $G$, by the projection maps $M \to X$ and $M \to G$.) The Kaluza-Klein metric on $M$ is defined to be:

$$\gamma = g_{JK}(\beta^J \otimes \beta^K + v^J \otimes v^K - \hat{v}^J \otimes \hat{v}^K) \tag{2.5}$$

where the two tetrads of one-forms $v^K$ and $\hat{v}^K$ are defined on $M$ by:

$$v^K = \alpha^K - |\sigma|\beta^K$$
$$\hat{v}^K = \hat{\alpha}^K - \omega^K_J \beta^J \tag{2.6}$$

where $\sigma$ is a complex dimensionless scalar field and $\omega^K_J$ is a matrix of real dimensionless scalar fields on $M$. The Kaluza-Klein metric $\gamma$ depends only on the fields $\beta^K$, $\sigma$, and $\omega^K_J$ for its dynamics, since ($\alpha^K$, $\hat{\alpha}^K$) is the fixed basis of one-forms chosen for the dual of the Lie algebra of $G$. Note from formulas (2.5) and (2.6) that the Kaluza-Klein metric $\gamma$ does not depend on any physical constants, such as Newton's constant or electric charge. For vector fields v and w on $M$, we will denote the inner product with respect to the metric $\gamma$ by $<v, w>$.

Associated with the metric $\gamma$ is an orthonormal basis of vector fields $(v_K, a_K, \hat{a}_K)$ that are dual to the one-forms $(\beta^K, v^K, \hat{v}^K)$ on $M = X \times G$. With respect to the basis $(v_K, a_K, \hat{a}_K)$, the Kaluza-Klein metric $\gamma$ in formula (2.5) becomes:

$$\gamma = \begin{bmatrix} g_{JK} & 0 & 0 \\ 0 & g_{JK} & 0 \\ 0 & 0 & -g_{JK} \end{bmatrix} \tag{2.7}$$

Note that the vector fields $v_K$ are horizontal vector fields that are orthogonal to the vertical vector fields $(a_K, \hat{a}_K)$ on the Kaluza-Klein manifold $M = X \times G$.

Our goal in this section is to derive the usual Einstein-Maxwell-Dirac Lagrangian from the following Lagrangian for the dimensionless field components ($\beta_\alpha^K$, $\sigma$, $\omega_J^K$):

$$L = \frac{1}{2} R_v + \overline{v_K(\sigma+\varphi)} \, v^K(\sigma+\varphi) \tag{2.8}$$

where $R_v$ is the sum of sectional curvatures over the four-dimensional horizontal subspaces spanned by the orthonormal tetrad $v_K$ in each tangent space of $M$:

$$R_v = g^{JL} g^{KM} < R(v_J, v_K) v_L, v_M > \tag{2.9}$$

where $R(\,,\,)$ is the curvature two-form [40] associated with the Kaluza-Klein metric $\gamma$, and the dimensionless mass parameter $\varphi$ in the Kaluza-Klein Lagrangian $L$ acts as a complex scalar Higgs field on $M$ that generates the fermion mass $m_0$.

The Kaluza-Klein manifold $M = X \times G$ has a natural right action of $G$ defined by $h(x, g) = (x, gh)$ for each $(x, g) \in M$ and $h \in G$, which allows for dimensional reduction of the Kaluza-Klein equations [1] – [10]. For Kaluza-Klein solutions of interest, the Kaluza-Klein metric $\gamma$ is right invariant [9], [10]. For $\gamma$ to be right invariant, it is necessary and sufficient that $\beta^K$, $|\sigma|$, and $\omega_J^K$ which occur in the components of $\gamma$ depend only on the space-time points $x \in X$. Since $\zeta = |\sigma|$ depends only on $x \in X$, we consider Kaluza-Klein solutions for which the complex scalar field $\sigma : M \to C$, where $C$ is a one-dimensional complex vector space, has the following equivariant form [10]:

$$\sigma(x, y, \hat{y}) = \zeta(x) \, e^{i(y/\delta_0 + \hat{y}/\hat{\delta}_0)} \tag{2.10}$$

where $y$ and $\hat{y}$ are global coordinates of the two $U(1)$ factors of the gauge group $G$ generated by $a_3 = -\partial/\partial y$ and $\hat{a}_3 = -\partial/\partial \hat{y}$. The complex scalar field $\sigma: M \to C$ in formula (2.10) commutes with the action of $G$ on $M$ and $C$, as required for the equivariant scalar fields defined on the principal $G$-bundle $M = X \times G$ in Yang-Mills theory [39]. Similarly, we consider a dimensionless mass parameter $\varphi$ having the following equivariant form [10]:

$$\varphi(y, \hat{y}) = \varphi_0 \, e^{i(y/\delta_0 + \hat{y}/\hat{\delta}_0)} \qquad (2.11)$$

where $\varphi_0$ is a dimensionless constant. To derive this as a solution from a Kaluza-Klein Lagrangian, we subtract a quartic Higgs potential containing $\varphi$ from the Lagrangian $L$ of the form [12]:

$$Q(\varphi) = Q_0 \left( \left| \frac{\varphi}{\varphi_0} \right|^2 - 1 \right)^2 \qquad (2.12)$$

where $Q_0$ can be any sufficiently large positive constant having units of curvature.

The complex scalar fields $\sigma: M \to C$ and $\varphi: M \to C$ in formulas (2.10) and (2.11) commute with the action of $G$ on $M$ and $C$, as required for all scalar fields defined on the principal $G$-bundle $M = X \times G$ in Yang-Mills theory [39]. It can be shown that right invariant metrics and equivariant scalar fields on $M = X \times G$ are solutions of the Kaluza-Klein equations, which allow dimensional reduction. Restricted to these solutions, Kaluza-Klein theory reduces to Yang-Mills theory [9], [10].

Let $d\gamma$ denote the volume form on $M = X \times G$ defined by the Kaluza-Klein metric $\gamma$. (We do not confuse the symbol "$d$" with exterior differentiation since the metric $\gamma$ is not a differential form.) Similarly let $d\alpha$ and $d\beta$ denote the volume forms defined by the metrics $\alpha$ and $\beta$ on the manifolds $G$ and $X$, respectively. Since the one-forms $(\beta^K, \nu^K, \hat{\nu}^K)$ are orthonormal, we see from formulas (2.3), (2.5), and (2.6) that

$$d\gamma = d\beta \wedge d\alpha \qquad (2.13)$$

Therefore, the action for the dimensionless field components $(\beta_\alpha^K, \sigma, \omega_J^K)$ associated with the Lagrangian (2.8) is given by

$$S = \frac{1}{8\pi\kappa_0} \int L(\beta_\alpha^K, \sigma, \omega_J^K) \, d\beta \wedge d\alpha \qquad (2.14)$$

where $\kappa_0$ is Newton's constant.

**THEOREM:** The Kaluza-Klein Lagrangian $L$ in the action (2.14) has no explicit dependence on any physical constants other than for generating mass. Furthermore, the reduction of the Kaluza-Klein Lagrangian $L$ to precisely the usual Einstein-Maxwell-Dirac Lagrangian is accomplished in five steps as follows:

(1) Define the ratio of the fermion radii $\delta_0/\delta$ to equal $2/3$.
(2) Scale the dimensionless fields $\omega_J^K$, $\sigma$, and $\varphi$ by the factors $\kappa_0^{-1/2}$, $(\kappa_0\delta)^{-1/3}$, and $\delta^{-1}$, respectively, whose units are mass. (Note the very different mass scalings for the three fields.)
(3) Dimensionally reduce the Lagrangian $(8\pi\kappa_0)^{-1}L$ from the Kaluza-Klein manifold $M$ to the space-time $X$.
(4) Take the limit as $\delta$ goes to zero.
(5) Restrict the boson $SL(2,R) \times U(1)$ gauge group to the electromagnetic $U(1)$ subgroup.

**PROOF:** The horizontal vector fields $v_K$ in the Kaluza-Klein Lagrangian (2.8) have no explicit dependence on any physical constants. Hence, the Lagrangian (2.8) has no explicit dependence on any physical constants other than for generating mass. Note that the appearance of Newton's constant $\kappa_0$ dividing the total action in formula (2.14) plays no role classically, since it can be replaced by any constant of dimension length squared.

To prove the rest of the theorem, we first derive a local expression for the Lagrangian (2.8). Let $U \subset X$ be an open chart of the space-time $X$ on which coordinates denoted as $x^\alpha$ with $\alpha = 0, 1, 2, 3$ are defined. The gravitational field $\beta$ can then be expressed locally on $U \times G \subset M$ as follows:

$$\beta = g_{JK} \beta^J \otimes \beta^K = g_{\alpha\beta} dx^\alpha \otimes dx^\beta \tag{2.15}$$

where $dx^\alpha$ denote the one-forms on $U \times G$ induced by the projection map $U \times G \to U$ from the coordinate one-forms $dx^\alpha$ defined on $U$. Repeated space-time coordinate indices $\alpha, \beta, \gamma, \delta$ are summed from 0 to 3. Since $\beta^K = \beta^K_\alpha dx^\alpha$ and $b_K = b^\alpha_K \partial_\alpha$, where $(b_K, a_K, \hat{a}_K)$ are the vector fields dual to the one forms $(\beta^K, \alpha^K, \hat{\alpha}^K)$ and $(\partial_\alpha, a_K, \hat{a}_K)$ are the vector fields dual to the one forms $(dx^\alpha, \alpha^K, \hat{\alpha}^K)$ on $U \times G$, we obtain from formula (2.15):

$$g_{\alpha\beta} = g_{JK} \beta^J_\alpha \beta^K_\beta$$

$$g^{\alpha\beta} = g^{JK} b^\alpha_J b^\beta_K \tag{2.16}$$

where $g^{\alpha\beta}$ denotes the inverse of the metric tensor $g_{\alpha\beta}$. Note that like $g_{JK}$ and $g^{JK}$, the components $g_{\alpha\beta}$ and $g^{\alpha\beta}$ of the gravitational field are dimensionless.

Define constants $\kappa$, $\lambda$, $\hat{\lambda}$, $g$, $g_0$, $q$, and $q_0$ in terms of Newton's constant $\kappa_0$ and the higher dimensional radii $\delta$, $\delta_0$, $\hat{\delta}$, and $\hat{\delta}_0$ as follows:

$$\kappa = \frac{16\pi}{3}\kappa_0$$

$$\lambda = (\kappa\delta)^{1/3}, \qquad \hat{\lambda} = \kappa^{1/2}$$

$$g = \frac{(\kappa\delta)^{1/3}}{\delta}, \qquad 2q = \frac{\kappa^{1/2}}{\hat{\delta}} \qquad (2.17)$$

$$g_0 = \frac{(\kappa\delta)^{1/3}}{\delta_0}, \qquad 2q_0 = \frac{\kappa^{1/2}}{\hat{\delta}_0}$$

Then define the boson and fermion fields $(V_\alpha^K, F_\alpha^K, \rho)$ and mass parameter $\mu$ on $U \times G$ as follows:

$$\mu = \lambda^{-1}\varphi$$

$$\rho = \lambda^{-1}\sigma$$

$$F_\alpha^K = \lambda^{-1}|\sigma|\beta_\alpha^K \qquad (2.18)$$

$$V_\alpha^K = \hat{\lambda}^{-1}\omega_J^K \beta_\alpha^J$$

Note that by formulas (2.16) and (2.18) the fermion gauge potentials $F_\alpha^K$ satisfy the orthogonal constraint (1.2). Let $\delta_0 = \frac{2}{3}\delta$. Then, by formulas (2.17) and (2.18), the fermion mass in the Yang-Mills Lagrangian (1.3) is given by $m_0 = \frac{1}{2}g_0\mu = \frac{3}{4}\delta^{-1}\varphi$. Thus, apart from a numerical factor close to one, the dimensionless scalar field $\varphi$ is scaled by the mass factor $\delta^{-1}$ to obtain the fermion mass $m_0$. Similarly from formulas (2.17) and (2.18), apart from numerical factors close to one, the dimensionless fields $\sigma$ and $\omega_J^K$ are scaled by the mass factors $(\kappa_0\delta)^{-1/3}$ and $\kappa_0^{-1/2}$ to obtain $(F_\alpha^K, \rho)$ and $V_\alpha^K$, respectively. Note that this scaling of the dimensionless scalar fields $\sigma$ and $\varphi$ is performed prior to taking the limit in which $\delta$ goes to zero in Step (4), since these scalings depend on $\delta$.

Define a local coordinate tetrad $v_\alpha = \beta_\alpha^K v_K$ on $U \times G \subset M$ that spans the same four-dimensional horizontal distribution over the Kaluza-Klein manifold $U \times G$ as $v_K$. Using formula (2.18) we have:

$$v_\alpha = \partial_\alpha + \lambda F_\alpha^K a_K + \hat{\lambda} V_\alpha^K \hat{a}_K \qquad (2.19)$$

Then, substituting $v_K = b_K^\alpha v_\alpha$ into $R_v$ in formula (2.9) and using formula (2.16), the sum of sectional curvatures over the horizontal distribution spanned by $v_K$ becomes:

$$R_v = g^{\alpha\gamma} g^{\beta\delta} \langle R(v_\alpha, v_\beta) v_\gamma, v_\delta \rangle \qquad (2.20)$$

and using formulas (2.17) and (2.18), the Lagrangian (2.8) equals:

$$L = \frac{1}{2} R_v + \frac{8\pi\kappa_0}{g_0} \overline{v_\alpha(\rho+\mu)} v^\alpha(\rho+\mu) \qquad (2.21)$$

The sum of horizontal sectional curvatures $R_v$ in formula (2.20) is evaluated using the vector fields $(v_\alpha, a_K, \hat{a}_K)$ as a basis on $U \times G$. Note that with respect to this basis, the Kaluza-Klein metric (2.5) has the following components:

$$\gamma = \begin{bmatrix} g_{\alpha\beta} & 0 & 0 \\ 0 & g_{JK} & 0 \\ 0 & 0 & -g_{JK} \end{bmatrix} \qquad (2.22)$$

The local expressions of $v_\alpha$, $R_v$, and $\gamma$ given in formulas (2.19), (2.20), and (2.22) are equal to the usual expressions in Kaluza-Klein theory [1], [4]. That is, $\gamma$ is precisely the Kaluza-Klein metric for the gravitational field $g_{\alpha\beta}$ and the semi-direct product gauge potentials $(F_\alpha^K, V_\alpha^K)$.

We dimensionally reduce the Lagrangian $(8\pi\kappa_0)^{-1} L$ with $L$ given in formula (2.21) by substituting $F_\alpha^K$ and $V_\alpha^K$ depending only on space-time points $x \in X$, and also substituting the scalar fields $\rho = \lambda^{-1}\sigma$ and $\mu = \lambda^{-1}\varphi$ where the scalar fields $\sigma$ and $\varphi$ have the equivariant forms given in formulas (2.10) and (2.11), respectively. These substitutions commute with the Euler-Lagrange equations. Using these substitutions we first evaluate $v_\alpha$ in formula (2.19) on the scalar fields $\sigma$ and $\varphi$. We next obtain the commutators for the basis $(v_\alpha, a_K, \hat{a}_K)$ using formula (2.4), and then express $R_v$ in formula (2.20) using these commutators and $\gamma$ as in formula (2.22). Using formula (2.17) we define the constants $\lambda$, $\hat{\lambda}$, $g$, $g_0$, $q$, and $q_0$ in terms of Newton's constant $\kappa_0$ and the higher dimensional radii $\delta$, $\hat{\delta}$, $\delta_0$, and $\hat{\delta}_0$. Then with these substitutions, evaluation of $L$ as expressed in formula (2.21) shows that the dimensionally reduced Lagrangian $(8\pi\kappa_0)^{-1} L$ is exactly equal to the Einstein-Hilbert Lagrangian for the gravitational field $g_{\alpha\beta}$ plus the Yang-Mills Lagrangian $L_{YM}$ given in formula (1.3) for the boson and fermion fields $V_\alpha^K$, $F_\alpha^K$, and $\rho$. Furthermore, formula (2.17) gives:

$$\delta = \left(\frac{2q}{g^{3/2}}\right)\hat{\delta} \qquad (2.23)$$

Thus in the limit required to obtain Dirac's equation, that is, as the fermion coupling constant $g$ becomes infinitely large, the fermion radius $\delta$ must become vanishingly small compared to the boson radius $\hat{\delta}$. The same is true for the fermion radius $\delta_0 = (2/3)\delta$. Finally, to obtain the Einstein-Maxwell-Dirac Lagrangian we restrict the boson gauge potentials $V_\alpha^K$ to the U(1) electromagnetic gauge group by setting $V_\alpha^1 = V_\alpha^2 = V_\alpha^3 = 0$, in which case only the electromagnetic gauge potential $V_\alpha = V_\alpha^0$ is non-vanishing. **Q.E.D.**

## 3. CONCLUDING REMARKS

Formulas (1.3) and (2.8) set three conditions for exact equality of Yang-Mills and Kaluza-Klein Lagrangians. The first condition is that only horizontal vector fields, defined on the Kaluza-Klein manifold $M = X \times G$, occur in the Kaluza-Klein Lagrangian [5], [39]. The second condition is that the scalar curvature $R_v$ in the Kaluza-Klein Lagrangian be restricted to the horizontal subspace of each tangent space of M. The third condition is that there exist solutions derived from the Kaluza-Klein Lagrangian for which the scalar fields, defined on M, transform equivariantly under the gauge group [39]. Unlike previous theories that do not satisfy the requirement that every term in a Kaluza-Klein theory correspond to an exactly equivalent term in a Yang-Mills theory [5], the Kaluza-Klein Lagrangian (2.8), after dimensional reduction and multiplying by a constant, equals exactly the Yang-Mills Lagrangian (1.3) plus a term describing purely gravity.

By formulating the Kaluza-Klein Lagrangian (2.8) with the horizontal vector fields $v_K$, the orthogonal constraint (1.2) is eliminated. Furthermore, in the theorem of Section 2, we see that the horizontal vector fields $v_K$ have no explicit dependence on any physical constants. (Note that to accomplish this, the vertical vector fields in formula (2.4) are chosen differently than in previous work [10].) The higher dimensional radii of the Kaluza-Klein manifold show up in the unified Lagrangian (2.8) only implicitly when computing the horizontal scalar curvature $R_v$ through the commutators of the vector fields $v_K$. Note that the appearance of Newton's constant $\kappa_0$ dividing the total action in formula (2.14) plays no role in the classical field equations. From the proof of the theorem in Section 2, it is clear that the elimination of Newton's constant along with all coupling constants is only possible because the three constants $\lambda$, $\hat{\lambda}$, and $\kappa_1 = 8\pi\kappa_0/g_0$ in the Kaluza-Klein Lagrangian (2.19) – (2.21) are related by the relation $\kappa_1 = \lambda^2$. Note that the orthogonal constraint (1.2) determines that there can be only two independent constants $\lambda^{-1}$, $\hat{\lambda}^{-1}$ in formula (2.18), and not three. This relation is also expressed by the specific coefficients multiplying each term in the Yang-Mills Lagrangian (1.3). It is a nontrivial property of Dirac's bispinor equation that precisely these coefficients occur in the Yang-Mills Lagrangian (1.3).

Since Newton's constant can be eliminated from the classical Kaluza-Klein equations, what then is the physical origin of Newton's constant? The answer lies in quantum mechanics. Although the appearance of Newton's constant dividing the total action in formula (2.14) plays no role in the classical equations, in quantum mechanics, changing the action would affect the normalization of the fermion and boson fields. In quantum mechanics, solutions of Dirac's bispinor equation are normalized to be unit vectors in a Hilbert space with time as a parameter. This normalization of the bispinor fields imposes a non-classical normalization of the tensor fields representing the bispinor fields in the Kaluza-Klein Lagrangian (2.8). Since the transformation to the tensor fields involves Newton's constant and the higher dimensional radii, the normalization of the tensor fields depends on these constants. Thus, in lieu of Newton's constant, a fundamental set of constants can be chosen to be the higher dimensional radii and a quantum mechanical normalization constant. The higher dimensional radii are geometrical, whereas, the non-classical normalization of the tensor fields finds its origin in the Hilbert space formulation of quantum mechanics.

We will conclude with an application of the Kaluza-Klein tetrad Lagrangian (2.8) to a topic of current interest [41] – [47], which involves finding a unified theory of "distant parallelism" or teleparallel theory that would include bispinor fields. For example, a teleparallel form of Dirac's bispinor equation gives an alternative interpretation of neutron interferometry experiments viewed from a rotating frame in a Minkowski space-time [42]. Recently Vassiliev, modeling neutrinos with a teleparallel tetrad field, derived a tensor Lagrangian equivalent to the Weyl spinor Lagrangian [43]. According to Mosna and Pereira, seeking a unified teleparallel theory is motivated by the following argument: "Let M be a 4-manifold representing our physical space-time. It is well known that, if we want to introduce spinors in this context, the existence of a global moving frame (or tetrad) $\{e_a\}_{a=0}^3$ on M must be assumed. … The global basis of vector fields $\{e_a\}$ gives rise to both a Riemannian and a teleparallel structure on M." [46]

For the usual Einstein-Maxwell-Dirac equations, in a non-compact four-dimensional space-time with spinor structure, the teleparallel covariant derivatives can be defined on the Kaluza-Klein manifold $M = X \times G$ with respect to the global moving frame $(v_K, a_K, \hat{a}_K)$. The Lagrangian $L'$, which is the teleparallel equivalent of the unified Kaluza-Klein Lagrangian (2.8), is then given as follows.

Replacing the constant $\kappa$ with the constant $\kappa' = 3\kappa$ in formulas (2.17) – (2.20), it can be shown that the Lagrangian (2.21) equals up to a divergence:

$$L' = \frac{1}{2} T_v + \frac{1}{3} \overline{v_K(\sigma + \varphi)} \, v^K(\sigma + \varphi) \tag{3.1}$$

where

$$T_v = \frac{1}{4} <T_{JK}, T^{JK}> + \frac{1}{2} <T_{JK}, v_L><T^{LK}, v^J> - <T_{JK}, v^J><T^{LK}, v_L> \tag{3.2}$$

where $T_{JK} = T(v_J, v_K)$ is the teleparallel torsion tensor restricted to horizontal vector fields $v_K$. We see in formula (3.1) that the horizontal teleparallel scalar torsion $T_v$ replaces the horizontal Riemannian scalar curvature $R_v$ in the Kaluza-Klein Lagrangian (2.8). Note that unlike current theories [44] – [47] for which the teleparallel covariant derivatives are defined differently for gravitational, electromagnetic, and bispinor fields, the torsion tensor $T(,)$ is defined using a single definition of teleparallel covariant derivative on $M = X \times G$. It can be shown that the Lagrangian (3.1) reduces to the teleparallel equivalent of the Einstein-Hilbert Lagrangian for gravity [41], when both electromagnetic and bispinor fields vanish. Thus, the Lagrangian (3.1) extends the teleparallel equivalent of general relativity, for gravity alone, to the electromagnetic and bispinor fields, and is equal (up to a divergence) to the Kaluza-Klein Lagrangian (2.8).